\documentclass[11pt,a4paper]{article}
\pdfoutput=1

\usepackage[nosort]{cite}
\usepackage[bulletsep]{collref}

\usepackage[british]{babel} 

\usepackage{epsfig,rotating}
\usepackage{amsmath,amssymb,amsthm}
\usepackage{amsfonts}
\usepackage{mathrsfs}
\usepackage{bbm}
\usepackage{bm}

\usepackage{color}
\usepackage[textwidth = 430 pt, textheight = 630 pt]{geometry}
\definecolor{MyDarkBlue}{rgb}{0.15,0.25,0.45}

\usepackage[linktocpage=true]{hyperref}
\hypersetup{
colorlinks=true,
citecolor=MyDarkBlue,
linkcolor=MyDarkBlue,
urlcolor=MyDarkBlue,
pdfauthor={Christian S\"amann, Robert Wimmer, Martin Wolf},
pdftitle={A Twistor Description of Six-Dimensional N=(1,1) Super Yang-Mills Theory},
pdfsubject={hep-th}
breaklinks=true
}

\let\fn\footnote
\renewcommand{\footnote}[1]{\linespread{1.1}\fn{#1}\linespread{1.29}}

\makeatletter\renewcommand{\section}{\@startsection
{section}{1}{\z@}{-3.5ex plus -1ex minus
    -.2ex}{2.3ex plus .2ex}{\bf\mathversion{bold} }}
\makeatletter\renewcommand{\subsection}{\@startsection{subsection}{2}{\z@}{-3.25ex
plus -1ex minus
   -.2ex}{1.5ex plus .2ex}{\bf\mathversion{bold} }}
\makeatletter\renewcommand{\subsubsection}{\@startsection{subsubsection}{3}{-2.45ex}{-3.25ex
plus -1ex minus -.2ex}{1.5ex plus .2ex}{\it }}
\renewcommand{\thesection}{\arabic{section}}
\renewcommand{\thesubsection}{\arabic{section}.\arabic{subsection}}
\renewcommand{\@seccntformat}[1]{\@nameuse{the#1}.~~}

\renewcommand{\theequation}{\thesection.\arabic{equation}}
\makeatletter \@addtoreset{equation}{section}

\renewcommand*\l@section{\@dottedtocline{1}{0em}{2em}}
\renewcommand*\l@subsection{\@dottedtocline{2}{2em}{2.4em}}
\renewcommand*\l@subsubsection{\@dottedtocline{4}{3.8em}{3.7em}}

\renewcommand\tableofcontents{%
    \section*{\large\contentsname
        \@mkboth{%
          \MakeUppercase\contentsname}{\MakeUppercase\contentsname}}%
       {\baselineskip=15pt plus 2pt minus 1pt
    \@starttoc{toc}}%
}

\renewenvironment{thebibliography}[1]
     {\baselineskip=16pt plus 2pt minus 1pt
      \section*{\large\refname
        \@mkboth{\MakeUppercase\refname}{\MakeUppercase\refname}}%
     \list{\@biblabel{\@arabic\c@enumiv}}%
           {\settowidth\labelwidth{\@biblabel{#1}}%
            \leftmargin\labelwidth
            \advance\leftmargin\labelsep
            \@openbib@code
            \usecounter{enumiv}%
            \let\p@enumiv\@empty
            \renewcommand\theenumiv{\@arabic\c@enumiv}}%
      \sloppy
      \clubpenalty4000
      \@clubpenalty \clubpenalty
      \widowpenalty4000%
      \sfcode`\.\@m
 \catcode`\^^M=10%
}

\newcommand{\appendices}{
\section*{Appendix}\label{appendices}\setcounter{subsection}{0}
\addcontentsline{toc}{section}{Appendix}
\setcounter{equation}{0}
\makeatletter
\renewcommand{\theequation}{\Alph{subsection}.\arabic{equation}}
\renewcommand{\thesubsection}{\Alph{subsection}}
\@addtoreset{equation}{subsection}
\makeatother
}

\numberwithin{lemma}{section}

\numberwithin{definition}{section}

\numberwithin{theorem}{section}

\numberwithin{prop}{section}

\numberwithin{cor}{section}

\def\periodb#1{\setbox0=\hbox{$#1$}#1\hskip-\wd0\hbox to\wd0{-}}

\newcommand{\CN}{\mathcal{N}}
\newcommand{\CO}{\mathcal{O}}

\newcommand{\CE}{\mathcal{E}}

\newcommand{\frU}{\mathfrak{U}}

\newcommand{\FC}{\mathbbm{C}}     			

\newcommand{\PP}{{\mathbbm{P}}}    			

\newcommand{\dpar}{\partial}     			
\newcommand{\embd}{{\hookrightarrow}}     		
\newcommand{\eps}{{\varepsilon}}			

\newcommand{\ald}{{\dot{\alpha}}}     			
\newcommand{\md}{{\dot{m}}}     			
\newcommand{\nd}{{\dot{n}}}     			
\newcommand{\bed}{{\dot{\beta}}}

\newcommand{\eand}{{~~~\mbox{and}~~~}}     		

\newcommand{\der}[1]{\frac{\dpar}{\dpar #1}}   		

\newcommand{\agl}{\mathfrak{gl}}     			

\newcommand{\sSU}{\mathsf{SU}}
\newcommand{\sSL}{\mathsf{SL}}
\newcommand{\sGL}{\mathsf{GL}}

\newcommand{\sSpin}{\mathsf{Spin}}

\newcommand{\remark}[1]{}     				
\def\tyng(#1){\hbox{\tiny$\yng(#1)$}}			
\def\tyoung(#1){\hbox{\tiny$\young(#1)$}}			

\begin{document}
\begin{titlepage}

\setcounter{page}{0}
\renewcommand{\thefootnote}{\fnsymbol{footnote}}

\begin{flushright}
 EMPG--12--01\\ HWM--12--01\\ DMUS--MP--12/01
\end{flushright}

\begin{center}

{\LARGE\textbf{\mathversion{bold}A Twistor Description of Six-Dimensional\\ $\CN=(1,1)$ Super Yang--Mills Theory}\par}

\vspace{1cm}

{\large
Christian S\"amann$^{a}$, Robert Wimmer$^{b}$, and Martin Wolf$^{c}$
\footnote{{\it E-mail addresses:\/}
\href{mailto:c.saemann@hw.ac.uk}{\ttfamily c.saemann@hw.ac.uk}, \href{robert.wimmer@ens-lyon.fr}{\ttfamily robert.wimmer@ens-lyon.fr},
\href{mailto:m.wolf@surrey.ac.uk}{\ttfamily m.wolf@surrey.ac.uk}
}}

\vspace{1cm}

{\it 
$^a$ Maxwell Institute for Mathematical Sciences\\
Department of Mathematics,
Heriot--Watt University\\
Edinburgh EH14 4AS, United Kingdom\\[.5cm]

$^b$
Universit\'e de Lyon, Laboratoire de Physique, UMR 5672, CNRS\\
\'Ecole Normale Sup\'erieure de Lyon\\
46, all\'ee d'Italie, F-69364 Lyon cedex 07, France\\[.5cm]

$^c$
Department of Mathematics,
University of Surrey\\
Guildford GU2 7XH, United Kingdom

}

\vspace{1cm}

{\bf Abstract}
\end{center}
\vspace{-.3cm}

\begin{quote}
We present a twistor space that describes super null-lines on six-dimensional $\CN=(1,1)$ superspace. We then show that there is a one-to-one correspondence between holomorphic vector bundles over this twistor space and solutions to the field equations  of $\CN=(1,1)$ super Yang--Mills theory. Our constructions naturally reduce to those of the twistorial description of maximally supersymmetric Yang--Mills theory in four dimensions.
\vfill
\noindent 15th May 2012 

\end{quote}

\setcounter{footnote}{0}\renewcommand{\thefootnote}{\arabic{thefootnote}}

\end{titlepage}

\tableofcontents 

\bigskip
\bigskip
\hrule
\bigskip
\bigskip

\section{Introduction and results}

The twistor description of solutions to chiral field equations in six dimensions goes back to the work of Hughston \cite{Hughston:1987he}. For recent works in this direction, see also \cite{Saemann:2011nb,Mason:2011nw} and references therein. A corresponding twistor description of solutions to non-chiral field equations in six dimensions, such as the equations of motion of Yang--Mills theory with maximal $\CN=(1,1)$ supersymmetry, has only been developed partially \cite{Devchand:1985au,Harnad:1987xq,Harnad:1995zy}.

The purpose of this letter is to give a complete twistor description of the maximally or $\CN=(1,1)$ supersymmetric Yang--Mills (MSYM) equations in six dimensions with an emphasis on the underlying geometries.\footnote{Notice that twistor methods have recently been applied in the description of scattering amplitudes in this theory, see e.g.~\cite{Cheung:2009dc,Dennen:2009vk,Brandhuber:2010mm,Dennen:2010dh}. Our way of describing ambitwistor space might prove useful in this context.} It is known that these equations can be encoded in constraint equations for a connection on superspace \cite{Harnad:1985bc}, which in turn correspond to the integrability condition of this connection along super null-lines \cite{Devchand:1985au,Samtleben:2009ts}. We start by describing a twistor correspondence for null-lines in six-dimensional space-time in some detail. We then present the corresponding supersymmetric extension for maximal $\CN=(1,1)$ supersymmetry. The resulting twistor space, denoted by $L^{9|8}$, turns out to be a rank-$5|8$ holomorphic supervector bundle over the four-dimensional Gra\ss mannian $G_{2,4}$. Next, we derive a Penrose--Ward transform to establish a one-to-one correspondence between equivalence classes of certain holomorphic vector bundles over $L^{9|8}$ and gauge equivalence classes of solutions to the equations of motion of six-dimensional MSYM theory. We end by demonstrating how our constructions reduce to those appearing in the twistorial description of MSYM theory in four dimensions \cite{Witten:1978xx,Isenberg:1978kk,Isenberg:1978qd}. 

Throughout this letter, we shall be working in the complex setting. Concretely, our six-dimensional space-time is a copy of $\FC^6$. If desired, however, reality conditions can be imposed at any point of our constructions, cf.\  \cite{Saemann:2011nb,Mason:2011nw}.

\section{Ambitwistor space $L^{9|8}$ of $\CN=(1,1)$ superspace}

In this section, we shall construct an ambidextrous twistor space (or ambitwistor space for short) $L^{9|8}$ of  six-dimensional $\CN=(1,1)$ superspace. This twistor space is very similar in spirit to the ambitwistor space of four-dimensional $\CN=3$ superspace \cite{Witten:1978xx,Isenberg:1978kk,Isenberg:1978qd}: while the latter parametrises super null-lines in four dimensions, $L^{9|8}$ parametrises certain super null-lines in six dimensions.  We shall first describe the body $L^9$ of the supermanifold $L^{9|8}$ in detail, before we come to the supersymmetric extension. Our notation and conventions are close to those of \cite{Saemann:2011nb}.

\subsection{Construction of the body $L^9$ of $L^{9|8}$}

\paragraph{Outline of the construction.} As usual in twistor geometry, we would like to establish a double fibration in which a correspondence space is simultaneously fibred over both twistor space and complexified flat space-time $\FC^6$. The correspondence space in such a twistor fibration is a direct product of two manifolds.\footnote{If we are considering a compactified space-time, this direct product has to be compactified appropriately.} The first factor in this product  is space-time itself. The second factor is the moduli space of linear subspaces of space-time that we wish to describe with the twistor correspondence, restricted to those through the origin. Note that this makes the correspondence space the space of such linear subspaces with a given base point. Different base points may describe the same subspace, and modding out the dependence on equivalent base points, we obtain twistor space. In this letter, we are interested in light-rays or null-lines in six dimensions. We shall see below that the space of null-lines through the origin is given by the four-dimensional Gra\ss mannian $G_{2,4}$, which is the space of two-planes in $\FC^4$. The correspondence space, which we shall denote by $F^{10}$, is therefore ten-dimensional and we have $F^{10}\cong\FC^6\times G_{2,4}$. Modding out the dependence on equivalent base points amounts to quotenting the correspondence space by an (integrable) rank-one distribution known as a twistor distribution. This yields a nine-dimensional complex manifold which we denote by $L^9$. Altogether, we have the following double fibration:
\begin{equation}\label{eq:DoubleFibrationBos}
 \begin{picture}(50,40)
  \put(0.0,0.0){\makebox(0,0)[c]{$L^{9}$}}
  \put(64.0,0.0){\makebox(0,0)[c]{$M^{6}$}}
  \put(34.0,33.0){\makebox(0,0)[c]{$F^{10}$}}
  \put(7.0,18.0){\makebox(0,0)[c]{$\pi_1$}}
  \put(55.0,18.0){\makebox(0,0)[c]{$\pi_2$}}
  \put(25.0,25.0){\vector(-1,-1){18}}
  \put(37.0,25.0){\vector(1,-1){18}}
 \end{picture}
\end{equation}
Here, the projection $\pi_1$ is the quotient map by the distribution and $\pi_2$ is the trivial projection. In the following, we shall discuss this double fibration, and in particular the structure of the space $L^9$, in more detail.

\paragraph{Null-lines in six dimensions.} 
For simplicity, we shall work in spinor notation on $M^6\cong\FC^6$, that is, we identify the tangent bundle $T_{M^6}$ with the antisymmetric tensor product $S\wedge S$ of the rank-four bundle of anti-chiral spinors $S$ over $M^6$. Correspondingly, we shall use local coordinates $x^{AB}=-x^{BA}$ with $A,B,\ldots=1,\ldots,4$ and take the (flat) metric $g_{AB,CD}:=\frac12\eps_{ABCD}$, where $\eps_{ABCD}$ is the Levi-Civita symbol in four dimensions. 

A null-vector $\lambda^{AB}$ in $M^6$ satisfies the equation
\begin{equation}\label{eq:Pluecker}
 \tfrac12\eps_{ABCD}\lambda^{AB}\lambda^{CD}\ =\ 0~.
\end{equation}
Null-lines are then obtained from null-vectors via the identification $\lambda^{AB}\sim \varrho\lambda^{AB}$ with $\varrho\in \FC^*:=\FC\setminus\{0\}$. The resulting equivalences classes describe points on the Gra\ss mannian $G_{2,4}$, the homogeneous coordinates $\lambda^{AB}$ are called Pl\"ucker coordinates, and \eqref{eq:Pluecker} is called the Pl\"ucker relation. The space $G_{2,4}$ features prominently in four-dimensional twistor correspondences, and a detailed account can be found, e.g., in \cite{Ward:1990vs}. In the following, we merely recall a few facts necessary for our discussion. 

Pl\"ucker coordinates provide an embedding of $G_{2,4}$ into $\PP^5$ via the quadric \eqref{eq:Pluecker} with the $\lambda^{AB}$ being the six homogeneous coordinates on $\PP^5$. Furthermore, as a coset space, the Gra\ss mannian $G_{2,4}$ is given by 
\begin{equation}
 G_{2,4}\  \cong\ \frac{\sSL(4,\FC)}{\sSL(2,\FC)\times \widetilde{\sSL(2,\FC)}}~,
\end{equation}
where $\sSL(2,\FC)\times \widetilde{\sSL(2,\FC)}$ is the little group of a null vector in $\FC^6$. The relation \eqref{eq:Pluecker} implies that the Pl\"ucker coordinates factorise according to
\begin{equation}\label{eq:PlueckerFactorisation}
 \lambda^{AB}\ =\ \tfrac12\eps^{ABCD}\lambda_{Ca}\lambda_{Db}\,\eps^{ab}~,
\end{equation}
where $a,b=1,2$ and $\eps^{ab}$ is the invariant tensor for $\sSL(2,\FC)$ with $\eps_{ac}\eps^{cb}=\delta_a^b$. We therefore have homogeneous coordinates $(\lambda_{Aa})\in {\rm Mat}_{4\times 2}(\FC)$ and a coset description
\begin{equation}
 G_{2,4}\ \cong \ \frac{{\rm Mat}_{4\times 2}(\FC)}{\sSL(2,\FC)\times\FC^*}~.
\end{equation}

Every plane $\lambda$ in $\FC^4$ has a natural dual $\mu$, which is spanned by a pair of chiral spinors $\mu^{A\dot a}$ for $\dot a,\dot b,\ldots=1,2$ with $\lambda_{Aa}\mu^{A\dot a}=0$. The $\mu^{A\dot a}$ represent homogeneous coordinates on a dual Gra{\ss}mannian $\tilde G_{2,4}$ and they define a set of dual Pl\"ucker coordinates $\mu_{AB}$ according to
\begin{equation}
 \mu_{AB}\ =\ \tfrac12\eps_{ABCD}\mu^{C\dot c}\mu^{D\dot d}\eps_{\dot c\dot d}~.
 \end{equation}
The indices $\dot a,\dot b,\ldots=1,2$ are to be understood as indices of the subgroup $\widetilde{\sSL(2,\FC)}$ of the little group and $\eps_{\dot a\dot b}$ is the invariant tensor of $\widetilde{\sSL(2,\FC)}$ with $\eps_{\dot a\dot c}\eps^{\dot c\dot b}=\delta_{\dot a}^{\dot b}$. Furthermore, the two Gra{\ss}mannians $G_{2,4}$ and $\tilde G_{2,4}$ can be identified via 
\begin{equation}\label{eq:GrassIden}
\lambda^{AB}\ =\ \tfrac{1}{2}\eps^{ABCD}\mu_{CD}\quad\Longleftrightarrow\quad
\tfrac12\eps^{ABCD}\lambda_{Ca}\lambda_{Db}\,\eps^{ab}\ =\ \mu^{A\dot a}\mu^{B\dot b}\eps_{\dot a\dot b}~.
\end{equation}
The above equality, like all equalities in the following involving homogeneous coordinates, is to be understood as an equality of equivalence classes.

\paragraph{Double fibration.} 
So far, we have seen that the correspondence space $F^{10}$ is topologically $\FC^6\times G_{2,4}$ and it is trivially fibred over space-time. We may coordinatise $F^{10}$ by either $(x^{AB},\lambda_{Aa})$ or $(x^{AB},\lambda_{AB})$. To mod out the dependence of the null-lines on equivalent base points, we quotient the correspondence space by the rank-one twistor distribution that is generated by the vector field
\begin{equation}\label{eq:DistributionL9}
 V\ :=\ \lambda^{AB}\der{x^{AB}}\ =\ \tfrac12\eps^{ABCD}\lambda_{Ca}\lambda_{Db}\,\eps^{ab}\der{x^{AB}}~.
\end{equation} 
The resulting space is a nine-dimensional complex manifold $L^9$. Note that by construction, the twistor space $L^9$ is a rank-five holomorphic vector bundle over $ G_{2,4}$. Let us now give a few more details about the geometry of $L^9$.

To this end, consider the dual tautological bundle\footnote{For more details on Gra\ss mannians and bundles over them, see e.g.\ \cite{Manin:1988ds}.} $T^\vee$ over $G_{2,4}$. This rank-two holomorphic vector bundle is generated by its global sections. We can parametrise the latter by moduli $r^A\in \FC^4\cong S$ according to $v_a=r^A\lambda_{Aa}$, where $S$ is the anti-chiral spin bundle over space-time. This allows us to write down the following short exact sequence:
\begin{equation}\label{eq:sesCE}
 0\ \longrightarrow \ E \ \longrightarrow\ \FC^4\otimes T^\vee \ \xrightarrow{\kappa\;:\;v^A_a\mapsto v^A_{(a}\lambda_{Ab)}} \ \odot^2 T^\vee \ \longrightarrow\ 0 ~.
\end{equation}
Notice that $\kappa$ has rank three such that $E$ is a rank-five holomorphic vector bundle, and its sections obey
\begin{equation}\label{eq:IntersectionCondition}
 v^A_{(a}\lambda_{Ab)}\ =\ 0~.
\end{equation}
The short exact sequence \eqref{eq:sesCE} induces a long exact sequence of cohomology groups and since all higher cohomology groups of $T^\vee$ and $\odot^2T^\vee$ vanish, we conclude that $H^0(G_{2,4},E)\cong \FC^6$ and $H^q(G_{2,4},E)=0$ for $q\geq1$. Global holomorphic sections of $E$ are of the form $v^A_a=p^{AB}\lambda_{Ba}$ with $p^{AB}\in \wedge^2 S\cong \FC^6$, and $E$ is generated by these sections.

In fact, we can identify $E$ with the twistor space $L^9$ provided we identify the moduli $p^{AB}$ with the space-time coordinates $x^{AB}$: the projection $\pi_1:F^{10}\to L^9$ is given by $\pi_1:(x^{AB},\lambda_{Aa})\mapsto(v^A_a,\lambda_{Aa})$ with
\begin{equation}\label{eq:IncidenceBosonic}
 v^A_a\ =\ x^{AB}\lambda_{Ba}~,
\end{equation}
and the vector fields \eqref{eq:DistributionL9} generating the twistor distribution indeed annihilate the $v^A_a$. We shall refer to the relation \eqref{eq:IncidenceBosonic} as the incidence relation. This relation implies a geometric twistor correspondence: points $(v,\lambda)$ in $L^9$ correspond to null-lines $\ell_{(v,\lambda)}=\pi_2(\pi_1^{-1}(v,\lambda))$ in space-time given by $x^{AB}=x_0^{AB}+\tau \lambda^{AB}$, where $x_0^{AB}$ is a particular solution to the incidence relation \eqref{eq:IncidenceBosonic} and $\tau\in\FC$. Vice versa, points $x$ in space-time correspond to submanifolds $\hat x=\pi_1(\pi_2^{-1}(x))\hookrightarrow L^9$ bi-holomorphic to the Gra{\ss}mannian $G_{2,4}$. Note that the above null-lines are the null-lines obtained by intersecting two three-planes which are totally null (so-called $\alpha$-planes) \cite{Hughston:1986hb}. Note also that we have just derived twistor space from space-time. Inversely, one can derive space-time from twistor space using Kodaira's theorem of 
relative deformation theory: because of $H^1(G_{2,4},L^9)=0$, there are no obstructions to relative deformations of $G_{2,4}$ inside $L^9$. Thus, we have a family of such deformations whose moduli space $H^0(G_{2,4},L^9)$ can indeed be identified with space-time $\FC^6$.

\paragraph{Other descriptions.}
In addition to using $G_{2,4}$, we may also use the dual Gra{\ss}mannian $\tilde G_{2,4}$ with homogeneous coordinates $\mu^{A\dot a}$ (or Pl\"ucker coordinates $\mu_{AB}$). Equipping (dual) space-time $\tilde M^6$ with coordinates $y_{AB}$, we can associate a correspondence space $\tilde F^{10}$ with coordinates $(y_{AB},\mu^{A\dot a})$ and introduce a twistor distribution generated by 
\begin{equation}
 \tilde V\ :=\ \mu_{AB}\der{y_{AB}}\ =\ \tfrac12\eps_{ABCD}\mu^{C\dot c}\mu^{D\dot d}\eps_{\dot c\dot d}\der{y_{AB}}~.
\end{equation} 
Altogether, we have a dual double fibration
\begin{equation}\label{eq:DoubleFibrationBosDual}
 \begin{picture}(50,40)
  \put(0.0,0.0){\makebox(0,0)[c]{$\tilde L^{9}$}}
  \put(64.0,0.0){\makebox(0,0)[c]{$\tilde M^{6}$}}
  \put(34.0,33.0){\makebox(0,0)[c]{$\tilde F^{10}$}}
  \put(7.0,18.0){\makebox(0,0)[c]{$\tilde \pi_1$}}
  \put(55.0,18.0){\makebox(0,0)[c]{$\tilde \pi_2$}}
  \put(25.0,25.0){\vector(-1,-1){18}}
  \put(37.0,25.0){\vector(1,-1){18}}
 \end{picture}
\end{equation}
where  $\tilde\pi_2$ is the trivial projection and $\tilde \pi_1:\tilde F^{10}\to \tilde L^9$ is given by $\tilde \pi_1:(y_{AB},\mu^{A\dot a})\ \mapsto\ (w_A^{\dot a},\mu^{A\dot a})$ with $w_A^{(\dot a}\mu^{A\dot b)}=0$ and  $w_A^{\dot a}=y_{AB}\mu^{B\dot a}$. This incidence relation yields an analogous geometric twistor correspondence to the one above: points in $\tilde{L}^9$ correspond to null-lines in $\tilde M^6$ and points in $\tilde M^6$ correspond to embedding of $\tilde G_{2,4}$ in $\tilde L^9$. Geometrically, the null-lines just described arise from intersecting two dual three-planes which are totally null  (so-called $\beta$-planes) \cite{Hughston:1986hb}.  Note that also $\tilde L^9$ can be described by a short exact sequence of the form \eqref{eq:sesCE}.

The manifolds $L^9$ and $\tilde{L}^9$ yield a description of null-lines in terms of chiral spinors $\lambda_{Aa}$ and anti-chiral spinors $\mu^{Aa}$, respectively. To obtain an ambidextrous description, that is, a description involving both  $\lambda_{Aa}$ and $\mu^{Aa}$ simultaneously, we  identify $G_{2,4}$ and $\tilde{G}_{2,4}$ via \eqref{eq:GrassIden} and write $\hat G_{2,4}$ and introduce $\hat{L}^9\hookrightarrow L^9\times \tilde{L}^9$ as the zero-locus
\begin{equation}\label{eq:QuadricBosonicAmbi}
\begin{aligned}
\big(v^A_av^B_b\eps^{ab}-\tfrac{1}{2}\eps^{ABCD}w_C^{\dot a}w_D^{\dot b}\eps_{\dot a\dot b}\big)\ {\rm mod}\ \lambda^{AB}\ =\ 0~.
\end{aligned}
\end{equation}
Because of \eqref{eq:QuadricBosonicAmbi}, global holomorphic sections are of the form 
\begin{equation}\label{eq:IncidenceBosonicAmbi}
 v^A_a\ =\ x^{AB}\lambda_{Ba}\eand w_A^{\dot a}\ =\ \tfrac12\eps_{ABCD}x^{CD}\mu^{B\dot a}~,
\end{equation}
and therefore $\hat L^9$ is a rank-five holomorphic vector bundle over $\hat G_{2,4}$.   Altogether, we have a double fibration
\begin{equation}\label{eq:DoubleFibrationBosAmbi}
 \begin{picture}(50,40)
  \put(0.0,0.0){\makebox(0,0)[c]{$\hat L^{9}$}}
  \put(64.0,0.0){\makebox(0,0)[c]{$M^{6}$}}
  \put(34.0,33.0){\makebox(0,0)[c]{$\hat F^{10}$}}
  \put(7.0,18.0){\makebox(0,0)[c]{$\pi_1$}}
  \put(55.0,18.0){\makebox(0,0)[c]{$\pi_2$}}
  \put(25.0,25.0){\vector(-1,-1){18}}
  \put(37.0,25.0){\vector(1,-1){18}}
 \end{picture}
\end{equation}
with the same space-time manifold $M^6$ as in \eqref{eq:DoubleFibrationBos} and
\begin{equation}
 \pi_1\, :\,(x^{AB},\{\lambda_{Aa},\mu^{A\dot a}\})\ \mapsto\ (\{v^A_a,w_A^{\dot a}\},\{\lambda_{Aa},\mu^{A\dot a}\})
 \end{equation}
is given by \eqref{eq:IncidenceBosonicAmbi}.

\paragraph{Double fibrations in Pl\"ucker coordinates.} 
In order to extend the above discussion to the supersymmetric setting with manifest maximal R-symmetry, we shall find it more convenient to work directly with Pl\"ucker coordinates. The advantage of these coordinates is that one can easily switch between chiral and anti-chiral descriptions by virtue of \eqref{eq:GrassIden}. The results of \cite{Harnad:1987xq,Harnad:1995zy} seem to suggest that a description using the homogeneous coordinates $\lambda_{Aa}$ and $\mu^{A\dot a}$ and having the full R-symmetry for $\CN=(1,1)$ supersymmetry manifest at the same time is not possible. Note, however, that in principle we can always substitute the Pl\"ucker coordinates by the homogeneous coordinates.

Recall that the Pl\"ucker coordinates define an embedding $i:G_{2,4}\embd \PP^5$ as the quadric \eqref{eq:Pluecker}. The bundle $\det T^\vee $ can be identified with $i^*\CO_{\PP^5}(1)$   \cite{Manin:1988ds}, and global sections of this bundle are given by $v=p^{AB}\lambda_{AB}$, $p^{AB}\in \FC^6$, and $\lambda_{AB}=\frac12\eps_{ABCD}\lambda^{CD}$. This bundle appears in the short exact sequence analogue to \eqref{eq:sesCE} involving Pl\"ucker coordinates,
\begin{equation}
 0\ \longrightarrow \ E \ \longrightarrow\  \FC^{16}\otimes \det T^\vee\ \xrightarrow{\kappa\;:\;v_B^A\mapsto (v_B^A\lambda^{BC}\,,\,v^A_{(B}\lambda_{C)A})} \ E' \ \longrightarrow 0~,
\end{equation}
where $\kappa$ has rank\footnote{To analyse the rank of such maps, it is helpful to consider them over a convenient point on the base manifold $G_{2,4}$, e.g. $\lambda_{12}\neq 0=\lambda_{13}=\ldots=\lambda_{34}$.} 11 and therefore $E$ has rank five. In more detail, the bundle $\FC^{16}\otimes \det T^\vee$ is generated by its global sections $v^A_B$ and we can write $v_B^A=p_B^A{}^{CD}\lambda_{CD}$. The eleven linear equations $v_B^A\lambda^{BC}=0$ and $v^A_{(B}\lambda_{C)A}=0$ in the fibre coordinates reduces the rank-16 bundle to a rank-five bundle $E$. In Pl\"ucker coordinates, global holomorphic sections of $E$ are of the form $v_B^A=p^{AC}\lambda_{CB}$ with $p^{AB}\in \wedge^2 S\cong \FC^6$. 

As before, identifying $L^9$ with $E$,  we can write the projection $\pi_1$ in the double fibration \eqref{eq:DoubleFibrationBos} as 
\begin{subequations}
\begin{equation}\label{eq:pi11}
 \pi_1:(x^{AB},\lambda^{AB})\ \mapsto\ (v_A^B,\lambda^{AB})~
\end{equation}
with an incidence relation of the form
\begin{equation}\label{eq:IncidenceBosonicPluck}
 v_B^A=x^{AC}\lambda_{CB}~.
\end{equation}
\end{subequations}
In a similar manner, we may repeat this analysis for $\tilde L^9$ and $\hat L^9$.

\subsection{Supersymmetric extension}

Let us now come to the $\CN=(1,1)$ supersymmetric extension of the twistor space $L^9$. We shall first construct twistor spaces for chiral and anti-chiral super null-lines, before extending these spaces to the twistor space of $\CN=(1,1)$ super null-lines. We shall use Pl\"ucker coordinates on all the Gra{\ss}mannians.

\paragraph{Twistor space for chiral super null-lines.} 
Let $\Pi$ be the Gra{\ss}mann parity changing operator. We start from $\CN=(1,0)$ superspace $M^{6|8}\cong \FC^{6|8}:=\FC^6\oplus\Pi\FC^8$, which we describe by Gra{\ss}mann even (bosonic) coordinates $x^{AB}\in \wedge^2S\cong \FC^6$ and Gra\ss mann odd (fermionic) coordinates $\theta^{mA}\in \FC^2\otimes \Pi S$. Here, the index $m=1,2$ is an index of $\sSL(2,\FC)$, the chiral subgroup of the R-symmetry group\footnote{not to be confused with the little group $\sSL(2,\FC)\times \widetilde{\sSL(2,\FC)}$} $\sSpin(4,\FC)\cong\sSL(2,\FC)\times \sSL(2,\FC)$. On $M^{6|8}$, we introduce the vector fields
\begin{equation}\label{eq:vecFieldsChiral}
 P_{AB}\ :=\ \der{x^{AB}}\eand D_{mA}\ :=\ \der{\theta^{mA}}+\eps_{mn}\theta^{nC}\der{x^{AC}}~,
\end{equation}
which satisfy the relation
\begin{equation}
\{D_{mA},D_{nB}\}\ =\ 2\eps_{mn}P_{AB}~.
\end{equation}
Chiral super null-lines are linear $1|4$-dimensional subspaces $\ell\hookrightarrow M^{6|8}$. The moduli space of such linear superspaces through the origin is still $G_{2,4}$ so that the correspondence space is $F^{10|8}\cong\FC^{6|8}\times G_{2,4}$. To obtain a twistor space, we have to mod out the dependence on equivalent base points. Here, this amounts to quotienting $F^{10|8}$ by a twistor distribution generated by the vector fields
\begin{equation}\label{eq:L94Distribution}
 V\ :=\ \lambda^{AB}P_{AB}\eand V_m^A\ :=\ \lambda^{AB}D_{mB}~.
\end{equation}
For each $\lambda_{AB}\in G_{2,4}$, we have four independent equations $\lambda_{AB}V^B_m=0$ and hence the twistor distribution is of rank-$1|4$. Moreover, it is integrable since $\{V_m^A,V_n^B\}=\eps_{mn}\lambda^{AB}V$ while $[V,V^A_m]=0$. Therefore, we have a foliation of $F^{10|8}$ by $9|8$-dimensional complex supermanifolds $L^{9|4}$.
By construction, $L^{9|4}$ is a rank-$5|4$ holomorphic supervector bundle over $G_{2,4}$, which we describe as a subbundle of $\FC^{16|8}\otimes \det T^\vee$:
\begin{equation}
 0\ \longrightarrow \ L^{9|4} \ \longrightarrow\  \FC^{16|8}\otimes \det T^\vee\ \stackrel{\kappa}{\longrightarrow} \ E' \ \longrightarrow\ 0~.
\end{equation}
Using coordinates $(v^A_B,\vartheta^{m}_A)$ in the fibres of $\FC^{16|8}\otimes \det T^\vee$ and the usual Pl\"ucker coordinates $\lambda^{AB}$ on the base, the map $\kappa$ is implicitly given by the relations
\begin{equation}\label{eq:FibreRelationsC}
 v_C^A\lambda^{BC}\ =\ 0~,\quad v_{(A}^C\lambda_{B)C}-\tfrac12\vartheta_A^m\vartheta_B^n\eps_{mn}\ = 0~,\eand \vartheta_{B}^m\lambda^{BA}\ =\ 0~.
\end{equation}
We can define a projection  $\pi_1: (x^{AB},\theta^{mA},\lambda^{AB})\mapsto (v^A_B,\vartheta_{A}^m,\lambda^{AB})$ with
\begin{equation}\label{eq:Projection1C}
 v^A_B\ =\ (x^{AC}-\tfrac{1}{2}\theta^{mA}\theta^{nC}\eps_{mn})\lambda_{CB}\eand
 \vartheta_{A}^m\ =\ \theta^{mB}\lambda_{BA}~,
\end{equation}
and the vector fields \eqref{eq:L94Distribution} generating the twistor distribution indeed annihilate $v^A_B$ and $\vartheta_A^m$. Equations \eqref{eq:Projection1C} represent a chiral super extension of the incidence relation \eqref{eq:IncidenceBosonicPluck}.

Because of the projection given in \eqref{eq:Projection1C} and the trivial projection $\pi_2:\FC^{6|8}\times G_{2,4}\rightarrow \FC^{6|8}$, we have the double fibration
\begin{equation}
 \begin{picture}(50,40)
  \put(0.0,0.0){\makebox(0,0)[c]{$L^{9|4}$}}
  \put(64.0,0.0){\makebox(0,0)[c]{$M^{6|8}$}}
  \put(34.0,33.0){\makebox(0,0)[c]{$F^{10|8}$}}
  \put(7.0,18.0){\makebox(0,0)[c]{$\pi_1$}}
  \put(55.0,18.0){\makebox(0,0)[c]{$\pi_2$}}
  \put(25.0,25.0){\vector(-1,-1){18}}
  \put(37.0,25.0){\vector(1,-1){18}}
 \end{picture}
\end{equation}
The geometric twistor correspondence here is between points on $L^{9|4}$ and chiral super null-lines in $M^{6|8}$ as well as between points on $M^{6|8}$ and holomorphic embeddings of $G_{2,4}$ into $L^{9|4}$. Explicitly, for any fixed point $(v^A_B,\vartheta_{A}^m,\lambda^{AB})\in L^{9|4}$, the incidence relation \eqref{eq:Projection1C} yields a ($1|4$-dimensional) chiral super null-line
\begin{equation}
  x^{AB}\ =\ x^{AB}_0+\tau\lambda^{AB}+\tau_C^m\lambda^{C[A}\theta^{nB]}_0\eps_{mn}\eand
  \theta^{mA}\ =\ \theta^{mA}_0 +\tau^m_B\lambda^{BA}~,
\end{equation}
where $(x^{AB}_0,\theta^{mA}_0)$ represent a particular solution to the incidence relation while $\tau$ and $\tau_A^m$ constitute one free bosonic parameter and four fermionic parameters (note that the matrix $\lambda^{AB}$ is of rank two, so only four out of the initial eight fermionic parameters enter). 

\paragraph{Twistor space for anti-chiral super null-lines.} 
A twistor space for anti-chiral super null-lines is constructed analogously. Here, we start from $\tilde{M}^{6|8}\cong \FC^{6|8}$, with bosonic coordinates $y_{AB}$ and fermionic coordinates $\theta_A^{\md}$, $\md=1,2$. The vector fields generating supertranslations read as 
\begin{equation}\label{eq:vecFieldsAChiral}
 \tilde P^{AB}\ :=\ \der{y_{AB}}\eand D^A_\md\ :=\ \der{\theta^\md_A}+\eps_{\md\nd}\theta^\nd_C\der{y_{AC}}~,
\end{equation}
and they satisfy the relation
\begin{equation}
\{D^A_\md,D^B_\nd\}\ =\ 2\eps_{\md\nd}\tilde P^{AB}~.
\end{equation}
The correspondence space is given by $\tilde F^{10|8}\cong \FC^{6|8}\times G_{2,4}$, and to mod out the dependence on equivalent base points, we have to quotient $\tilde F^{10|8}$ by the vector fields
\begin{equation}
 V\ :=\ \mu_{AB}\tilde P^{AB}\eand V_{\md A}\ :=\ \mu_{AB}D^B_\md~.
\end{equation}
The resulting vector bundle $\tilde{L}^{9|4}$ can be regarded as a subbundle of $\FC^{16|8}\otimes \det \tilde T^\vee$ over $\tilde G_{2,4}$, which we coordinatise by $(w_A^B,\vartheta^{\dot m A})$ in the fibres and the Pl\"ucker coordinates $\mu_{AB}$ on the base. The relations satisfied by the fibre coordinates are
\begin{equation}\label{eq:FibreRelationsAC}
 w^C_A\mu_{BC}\ =\ 0~,\quad w_C^{(A}\mu^{B)C}-\tfrac12\vartheta^{\md A}\vartheta^{\nd B}\eps_{\md\nd}\ =\ 0~,\eand \vartheta^{\dot m B}\mu_{BA}\ =\ 0~,
\end{equation}
and we have a double fibration 
\begin{equation}
 \begin{picture}(50,40)
  \put(0.0,0.0){\makebox(0,0)[c]{$\tilde L^{9|4}$}}
  \put(64.0,0.0){\makebox(0,0)[c]{$\tilde M^{6|8}$}}
  \put(34.0,33.0){\makebox(0,0)[c]{$\tilde F^{10|8}$}}
  \put(7.0,18.0){\makebox(0,0)[c]{$\pi_1$}}
  \put(55.0,18.0){\makebox(0,0)[c]{$\pi_2$}}
  \put(25.0,25.0){\vector(-1,-1){18}}
  \put(37.0,25.0){\vector(1,-1){18}}
 \end{picture}
\end{equation}
Here, the projection $\pi_1:(y_{AB},\theta^\md_A,\mu_{AB})\mapsto (w^A_B,\vartheta^{\dot m A},\mu_{AB})$ reads as
\begin{equation}\label{eq:Projection1AC}
w^A_B\ =\ (y_{BC}-\tfrac{1}{2}\theta_B^\md\theta_C^\nd\eps_{\md\nd})\mu^{CA}\eand \vartheta^{\md A}\ =\ \theta^{\md}_B\mu^{BA}~.
\end{equation}
This incidence relation yields again a geometric twistor correspondence between points  in $\tilde L^{9|4}$ and  ($1|4$-dimensional) anti-chiral super null-lines in $\tilde M^{6|4}$ as well as points in $\tilde M^{6|4}$ and submanifolds in $\tilde L^{9|4}$ bi-holomorphic to $\tilde G_{2,4}$.

\paragraph{Ambitwistor space.} 
Let us now come to the discussion of full $\CN=(1,1)$ supersymmetry. In particular, consider $\CN=(1,1)$ superspace $M^{6|16}\cong \FC^{6|16}$ equipped with coordinates $(x^{AB},\theta^{mA},\theta^\md_A)$. On this space, we have both the vector fields \eqref{eq:vecFieldsChiral} and \eqref{eq:vecFieldsAChiral} with the identification $\der{x^{AB}}=\tfrac12\eps_{ABCD} \der{y_{CD}}$. They generate the $\CN=(1,1)$ supersymmetry algebra in six dimensions,
\begin{equation}\label{eq:SUSYAlg}
\{D_{mA},D_{nB}\}\ =\ 2\eps_{mn}P_{AB}~,~~~\{D^A_\md,D^B_\nd\}\ =\ 2\eps_{\md\nd}P^{AB}~,~~~\{D_{mA},D^B_\nd\}\ =\ 0~,
\end{equation}
where $P^{AB}=\tfrac{1}{2}\eps^{ABCD}P_{CD}$. Note that here, the metric appears explicitly.

The correspondence space $F^{10|16}$ is then topologically $\FC^{6|16}\times G_{2,4}$ and coordinatised by $(x^{AB},\theta^{mA},\theta^\md_A,\lambda^{AB})$. On $F^{10|16}$, we introduce a rank-$1|8$ distribution generated by the vector fields
\begin{equation}\label{eq:GenTwistDist}
 V\ :=\ \lambda^{AB}P_{AB}\ =\ \lambda_{AB}P^{AB}~,\quad V_m^A\ :=\ \lambda^{AB}D_{mB}~,\eand V_{\md A}\ :=\ \lambda_{AB}D^B_\md~.
\end{equation}
This distribution is integrable, with the non-vanishing Lie brackets given by $\{V^A_m,V^B_n\}=\lambda^{AB}\varepsilon_{mn}V$ and $\{V_{\md A},V_{\nd B}\}=\varepsilon_{\md\nd}\lambda_{AB}V$. The quotient of $F^{10|16}$ by this distribution is the ambitwistor space $L^{9|8}$. It is a rank-$5|8$ supervector bundle over $G_{2,4}$ and its body is $L^9$. We describe $L^{9|8}$ as a subbundle of $\FC^{16|16}\otimes \det T^\vee$ with fibre coordinates $(v^A_B,\vartheta^m_A,\vartheta^{A\md})$ and the map $\kappa$ implicitly given by its kernel:
\begin{equation}\label{eq:Eqs-v-Super}
\begin{aligned}
v^{A}_C\lambda^{BC}+\tfrac12\vartheta^{\md A}\vartheta^{\nd B}\eps_{\dot m\dot n}\ &=\ 0~,&v_{(A}^C\lambda_{B)C}-\tfrac12\vartheta_A^m\vartheta_B^n\eps_{mn}\ &=\ 0~,\\
\lambda^{AB}\vartheta^m_B\ &=\ 0~,&\lambda_{AB}\vartheta^{\md B}\ &=\ 0~.
\end{aligned}
\end{equation} 
Note that $\kappa$ has indeed rank $11|8$, and we we have constructed a double fibration
\begin{equation}\label{eq:DoubleFibrationFull}
 \begin{picture}(50,40)
  \put(0.0,0.0){\makebox(0,0)[c]{$L^{9|8}$}}
  \put(64.0,0.0){\makebox(0,0)[c]{$M^{6|16}$}}
  \put(34.0,33.0){\makebox(0,0)[c]{$F^{10|16}$}}
  \put(7.0,18.0){\makebox(0,0)[c]{$\pi_1$}}
  \put(55.0,18.0){\makebox(0,0)[c]{$\pi_2$}}
  \put(25.0,25.0){\vector(-1,-1){18}}
  \put(37.0,25.0){\vector(1,-1){18}}
 \end{picture}
\end{equation}
As before, the projection $\pi_2$ is the trivial projection, while $\pi_1: (x^{AB},\theta^A_m,\theta^{\dot m}_A,\lambda^{AB})\mapsto (v^A_B,\vartheta_A^m,\vartheta^{A \dot m},\lambda^{AB})$ reads as
\begin{equation}\label{eq:IncidenceFullAmbi}
 \begin{aligned}
  v^A_B\ =\ (x^{AC}-\tfrac{1}{2}\theta^{mA}\theta^{nC}\eps_{mn})\lambda_{CB}+\tfrac12\theta_B^\md\theta_C^\nd\eps_{\md\nd}\lambda^{CA}~,\\
  \vartheta^{m}_A\ =\ \theta^{Bm}\lambda_{BA}~,\eand \vartheta^{A\dot m}\ =\ \theta^{\md }_B\lambda^{BA}~,\kern0.7cm
 \end{aligned}
\end{equation}
which describe global holomorphic sections of the bundle $L^{9|8}\to G_{2,4}$. 

The geometric twistor correspondence induced by the incidence relation \eqref{eq:IncidenceFullAmbi} is between points on $L^{9|8}$ and super null-lines in $M^{6|16}$ as well as between points on $M^{6|16}$ and holomorphic embeddings of $G_{2,4}$ into $L^{9|8}$. Explicitly, for a fixed point $(v,\vartheta,\lambda)\in L^{9|8}$, the above incidence relations determine a ($1|8$-dimensional) super null-line $\ell_{(v,\vartheta,\lambda)}\hookrightarrow M^{6|16}$ by
\begin{equation}
\begin{aligned}
  x^{AB}\ =\ x^{AB}_0+\tau\lambda^{AB}+\tau_C^m\lambda^{C[A}\theta^{nB]}_0\eps_{mn}+\tfrac12\eps^{ABCD}\tau^{\md E}\lambda_{E[C}\theta_0{}^{\nd}_{D]}\eps_{\md\nd}~,\\
  \theta^{mA}\ =\ \theta^{mA}_0 +\tau^m_B\lambda^{BA}~,\eand 
   \theta^{\md}_A\ =\ \theta_0{}^{\md}_A +\tau^{\md B}\lambda_{BA}~,\kern.7cm
  \end{aligned}
\end{equation}
where $(x^{AB}_0,\theta^{mA}_0,\theta_0{}^{\md}_{A})$ represent a particular solution to the incidence relation while $\tau$ and $(\tau_A^m,\tau^{\md A})$ constitute one free bosonic parameter and eight fermionic parameters. 

\section{Twistor construction of the MSYM equations in six dimensions}

\paragraph{Constraint equations.} We now come to the description of classical solutions to the equations of motion of MSYM theory on $M^6$ by means of holomorphic data on the ambitwistor space $L^{9|8}$. The key fact here is that these equations are equivalent to certain constraint equations for a connection on the superspace $M^{6|16}$ \cite{Harnad:1985bc} and furthermore, that these constraint equations  can in turn be interpreted as integrability conditions along certain null-lines \cite{Devchand:1985au,Samtleben:2009ts}. Concretely, the  equations of motion of MSYM theory in six dimensions are equivalent to the following set of constraint equations  \cite{Harnad:1985bc}:
\begin{equation}\label{eq:ConstraintSYM}
\begin{aligned}
\{\nabla_{mA},\nabla_{nB}\}\ &=\ 2\eps_{mn}\nabla_{AB}~,\\
\{\nabla_{mA},\nabla^B_\nd\}-\tfrac14\delta^A_B\{\nabla_{mC},\nabla^C_\nd\}\ &=\ 0~,\\
\{\nabla^A_\md,\nabla^B_\nd\}\ &=\ \eps_{\md\nd}\eps^{ABCD}\nabla_{CD}\ =\ 2\eps_{\md\nd}\nabla^{AB}~.
\end{aligned}
\end{equation}
Here, the covariant derivatives are given in terms of a gauge potential with components $A_{AB}$, $A_{mA}$ and $A^A_\md$ as 
\begin{equation}
 \nabla_{AB}\ :=\ \partial_{AB}+A_{AB}~,\quad
 \nabla_{mA}\ :=\ D_{mA} + A_{mA}~,\eand
 \nabla_\md^A\ :=\ D^A_\md + A^A_\md~,
\end{equation}
where the derivatives $\dpar_{AB}$, $D_{mA}$, and $D^A_\md$ were defined in \eqref{eq:vecFieldsChiral} and \eqref{eq:vecFieldsAChiral}. The component fields of MSYM theory appear as component fields in the superfield expansion of the gauge potential $(A_{AB},A_{mA},A^A_\md)$, cf.\ \cite{Harnad:1985bc}. As observed in \cite{Devchand:1985au,Samtleben:2009ts}, the constraint equations \eqref{eq:ConstraintSYM} can be understood as integrability conditions of an auxiliary linear system. In the notation of \cite{Samtleben:2009ts}, it reads as 
\begin{equation}\label{eq:AuxLinSys}
 \xi^{AB}\nabla_{AB} \psi\ =\ 0~,\quad
 \xi^{AB}\nabla_{mB} \psi\ =\ 0~,\eand
 \xi_{AB}\nabla_\md^{B}\psi\ =\ 0~,
\end{equation}
where the spectral parameter $\xi^{AB}$ is some six-vector that is null and therefore describes a point on $G_{2,4}$. Below, we shall see how this system arises from a Penrose--Ward transform.

\paragraph{Penrose--Ward transform.} 
In a general Penrose--Ward transform, one starts from an element $f$ of a cohomology group on a twistor space, which is the base of a correspondence space that is simultaneously fibred over a space-time. To perform the transform, one pulls $f$ back to the correspondence space and pushes it down to space-time. Here, we start from a rank-$r$ vector bundle $\CE$ over $L^{9|8}$, i.e.\ an element of the first \v Cech cohomology group $H^1(L^{9|8},\sGL(r))$, and transform it to a solution of the MSYM equations on $\FC^6$.

More explicitly, choose an open Stein covering ${\hat \frU}=\{\hat U_{a}\}$ of $L^{9|8}$ and let $f=\{f_{ab}\}$ on $\hat U_a\cap \hat U_b$ be the transition functions of $\CE$. We shall assume that $\CE$ becomes holomorphically trivial on any $\hat x =\pi_1(\pi_2^{-1}(x))\hookrightarrow L^{9|8}$.
Note that the leaves of the fibration $\pi_1$ in the double fibration \eqref{eq:DoubleFibrationFull} are topologically trivial, and therefore the cover $\hat \frU$ induces a cover $\frU'=\{U'_a\}$ with $U'_a=\pi_1^{-1}(U_a)$ on correspondence space $F^{10|16}$. The pull-back bundle $\CE':=\pi_1^*\CE$ can thus be described by transition functions $f'_{ab}$, which are pull-backs of the transition functions $f_{ab}$: $f'_{ab}=\pi_1^*f_{ab}$. Since we assumed that $\CE$ is holomorphically trivial on any $\hat x =\pi_1(\pi_2^{-1}(x))\hookrightarrow L^{9|8}$, the bundle $\CE'$ is holomorphically trivial on all of $F^{10|16}$. Therefore, we have a holomorphic splitting of $f'_{ab}$ according to
\begin{equation}
 f'_{ab}\ =\ (h'_a)^{-1}h'_b~,
\end{equation}
where the $h'_a$ are holomorphic functions on $U'_a$ taking values in $\sGL(r,\FC)$. 

By definition of the pull-back, the $f'_{ab}$ must be constant along the leaves of the fibration $\pi_1: F^{10|16}\to L^{9|8}$. Hence, they are annihilated by the vector fields \eqref{eq:GenTwistDist}, which implies
\begin{equation}\label{eq:preLinSys}
\begin{aligned}
 h'_a~V~ (h'_a)^{-1}\ &=\ h'_b~ V~ (h'_b)^{-1}~,\\
 h'_a~V^A_m~ (h'_a)^{-1}\ &=\ h'_b ~V^A_m~ (h'_b)^{-1}~,\\
 h'_a~V_{\md A}~ (h'_a)^{-1}\ &=\ h'_b ~V_{\md A}~ (h'_b)^{-1}~.
 \end{aligned}
\end{equation}
This allows us to introduce a globally defined relative differential one-form with components
\begin{equation}
\begin{aligned}
 A|_{U_a}\ &:=\ h'_a ~V~ (h'_a)^{-1}\ =:\ \lambda^{AB}A_{AB}~,\\
 A_m^A|_{U_a}\ &:=\ h'_a ~V^A_m~ (h'_a)^{-1}\ =:\ \lambda^{AB} A_{mB} ~,\\
 A^{\md A}|_{U_a}\ &:=\ h'_a ~V_{\md A}~  (h'_a)^{-1}\ =:\ \lambda_{AB}A_\md^{B}~.
 \end{aligned}
\end{equation}
Here, the components $A_{AB}$, $A_{mA}$, and $A_\md^{A}$ take values in $\agl(r,\FC)$ and depend only on space-time. The equations \eqref{eq:preLinSys} may thus be re-written as
\begin{equation}
 \lambda^{AB}\nabla_{AB} h'_a\ =\ 0~,\quad
 \lambda^{AB}\nabla_{mB} h'_a\ =\ 0~,\eand
 \lambda_{AB}\nabla_\md^{B}h'_a\ =\ 0~,
\end{equation}
which is equivalent to \eqref{eq:AuxLinSys}. From the rank-$r$ holomorphic vector bundle $\CE$, we thus constructed a gauge potential with components $A_{AB}$, $A_{mA}$, and $A_{\md}^A$ encoding a solution to the equations of motion of $\CN=(1,1)$ SYM theory in six dimensions with gauge group $\sGL(r,\FC)$. Requiring $\det\CE$ to be trivial amounts to reducing the gauge group $\sGL(r,\FC)$ to $\sSL(r,\FC)$. Moreover, appropriate reality condition may now be imposed to discuss the MSYM equations on Minkowski space-time and with gauge group $\sSU(r)$, see \cite{Saemann:2011nb,Mason:2011nw}  for details on various reality conditions.

Note that the Penrose transform is a map between equivalence classes of holomorphic vector bundles over $L^{9|8}$ becoming trivial on any $\hat{x}\cong G_{2,4}\embd L^{9|8}$ and gauge equivalence classes of solutions encoded in the gauge potential $(A_{AB},A_{mA},A_{\md}^A)$. Moreover, this map establishes a one-to-one correspondence between these equivalence classes.

\paragraph{Reduction to 4d.}
To close, let us briefly comment on how our twistor correspondence and Penrose--Ward transform reduce to those of MSYM theory in four dimensions. Recall that the space of super null-lines in $\CN=4$ superspace $M^{4|16}\cong\FC^{4|16}$ through the origin is given by $\PP^1\times \PP^1$, and thus the correspondence space is $F^{6|16}\cong \FC^{6|16}\times \PP^1\times \PP^1$. To obtain twistor space, one has to factor by an integrable rank-$1|8$ distribution, which yields a rank-$3|8$ vector bundle $L^{5|8}$ over $\PP^1\times \PP^1$. This bundle is a quadric in the space $\PP_\circ^{3|4}\times \PP_\circ^{3|4}$ with $\PP^{3|4}_\circ\cong \FC^{2|4}\otimes \CO_{\PP^1}(1)$. The corresponding double fibration reads as
\begin{equation}
 \begin{picture}(50,40)
  \put(0.0,0.0){\makebox(0,0)[c]{$L^{5|8}$}}
  \put(64.0,0.0){\makebox(0,0)[c]{$M^{4|16}$}}
  \put(34.0,33.0){\makebox(0,0)[c]{$F^{6|16}$}}
  \put(7.0,18.0){\makebox(0,0)[c]{$\pi_3$}}
  \put(55.0,18.0){\makebox(0,0)[c]{$\pi_4$}}
  \put(25.0,25.0){\vector(-1,-1){18}}
  \put(37.0,25.0){\vector(1,-1){18}}
 \end{picture}
\end{equation}

Explicitly, the reduction can be performed by splitting the spinor indices $A,B,\ldots$ of $\sSL(4,\FC)$ into spinor indices $(\alpha,\ald),(\beta,\bed),\ldots$ of $\sSL(2,\FC)\times \widetilde{\sSL(2,\FC)}$. The superspace $\FC^{6|16}$ with coordinates $(x^{AB},\theta^{mA},\theta^{\dot m}_A)$ is dimensionally reduced along the $x^{12}$- and $x^{34}$-directions to the superspace $\FC^{4|16}$ with coordinates $(x^{\alpha\ald},\theta^{i\alpha},\theta^\ald_i)$. To reduce $G_{2,4}$ to $\PP^1\times \PP^1$, we make the same reduction along Pl\"ucker coordinates $\lambda_{12}$ and $\lambda_{34}$. The remaining Pl\"ucker coordinates $\lambda_{\alpha\ald}$ can be factorised into the product of two sets $(\nu_\alpha,\tilde{\nu}_\ald)$ of homogeneous coordinates on $\PP^1\times \PP^1$: $\lambda_{\alpha\ald}=\nu_\alpha\tilde{\nu}_\ald$. By applying the analogous reductions to the vector fields of the twistor distribution, one obtains the twistor distribution determining $\pi_3$. This completes the reduction of the 
twistor correspondence. The Penrose--Ward transform now reduces in principle trivially, up to a technical issue: it is well-known that integrability along $1|8$-dimensional null-lines in $\FC^{4|16}$ yields the equations of MSYM theory in four dimensions up to an additional algebraic condition \cite{Witten:1978xx,Isenberg:1978kk,Isenberg:1978qd}. To circumvent this problems, one can reduce the manifest R-symmetry group in the formulation from $\sSL(4,\FC)$ to $\sSL(3,\FC)$ and thus study integrability along $1|6$-dimensional null-lines in $\FC^{4|12}$. This yields the $\CN=3$ SYM equations in four dimensions, which are equivalent to the MSYM equations. If we dimensionally reduce the constraint equations \eqref{eq:ConstraintSYM} of six-dimensional MSYM theory, we obtain both the constraint equations for four-dimensional MSYM theory as well as the algebraic condition.

\paragraph{Acknowledgements.}
We would like to thank Elizabeth Gasparim for discussions. C.S. was supported by an EPSRC Career Acceleration Fellowship. R.W. was supported in part by the F{\'e}d{\'e}ration de recherche A.M. Amp{\`e}re.

\bibliographystyle{latexeu}

\end{document}